\documentclass[pra,floatfix,showpacs,twocolumn,superscriptaddress]{revtex4}
\usepackage{amsmath}
\usepackage{latexsym}
\usepackage{bm}
\usepackage{amssymb}
\usepackage{graphics}
\usepackage{color}
\usepackage{amsthm}
\usepackage{bbm}

\begin{document}

\title[Single-experiment-detectable nonclassical correlation witness]{Single-experiment-detectable nonclassical correlation witness}

\author{Robabeh Rahimi}
\affiliation{Interdisciplinary Graduate School of Science and Engineering, Kinki University, 3-4-1 Kowakae, Higashi Osaka, Osaka 577-8502, Japan}
\affiliation{Departments of Chemistry and Materials Science, Graduate School of Science, Osaka City University, Osaka 558-8585, Japan}
\author{Akira SaiToh}
\affiliation{Interdisciplinary Graduate School of Science and Engineering, Kinki University, 3-4-1 Kowakae, Higashi Osaka, Osaka 577-8502, Japan}

\begin{abstract}
Recent progress in theories of quantum information has determined nonclassical correlation defined differently from widely-used entanglement as an important property to evaluate computation and communication with mixed quantum states. We introduce an operational method to detect nonclassical correlation of bipartite systems. In this method, we use particular maps analogous to the well-established entanglement witnesses. Thus, the maps are called nonclassical correlation witness maps. Furthermore, it is proved that such a map can be generally decomposed so that a single-run experiment is feasible for implementation in bulk-ensemble systems.
\end{abstract}


\maketitle

It has been a couple of decades since when entanglement has returned back to the center of interests and is being studied by several research groups. The main reason for such popularity has been under the general believe that entanglement should be one of the distinct resources of power for quantum information processing to surpass any corresponding classical counterpart. Recently, however, the statement has been revisited with further studies on nonclassical correlation defined differently from entanglement. In particular, the computational power beyond classical schemes that is achieved by employing highly mixed, hence separable, quantum states is known to be related to nonclassical correlation, other than entanglement, of quantum systems.
A typical case where a nonentangled but nonclassically correlated state plays a critical role is computation with an NMR \cite{NMRQC} or Electron Nuclear DOuble Resonance (ENDOR) \cite{ENDORQC} bulk ensemble quantum computer. These systems, at room temperature, operate without entanglement. Nevertheless, the inhabitant quantum state is a mixed state possessing nonclassical correlation, which exhibits features that are out of reach for truly classical states. For instance, in quantum superdense coding, the protocol produces appropriate results when Bob successfully decodes Alice's encoding. In NMR implementation of superdence coding, it is proved \cite{c-R06} that Bob extracts the expected outcomes as long as the output signals are detectable as to whether the polarization of each qubit is positive or negative and this is possible without entanglement. Another example is the approximate trace estimation of a unitary matrix in a deterministic quantum computing with a single (pseudo-) pure qubit (DQC1) \cite{c-KL98}. It is known \cite{c-D08} to involve a considerable increase in nonclassical correlation but not in entanglement.

Recall that for two systems, A and B, in the separability paradigm \cite{P96}, a state that can be prepared by local operations and classical communications (LOCC), namely a separable state, is
\[
 \rho_{\rm sep}=\sum_{i}w_{i}\rho^{\rm A}_{\rm i}\otimes \sigma^{\rm B}_{\rm i},
\]
where $w_i$'s are positive weights; $\rho_{i}^{\rm A}$'s and $\sigma_{i}^{\rm B}$'s are local states. An entangled state $\rho_{\rm ent}$ is an inseparable state.

In the Oppenheim-Horodecki paradigm \cite{c-OH02,c-H05}, a (properly) classically correlated state is
\[
 \rho_{\rm pcc}=\sum_{ij}e_{ij}|e_i\rangle^{\rm A}\langle e_i|\otimes |e_j\rangle^{\rm B}\langle e_j|,
\]
where $e_{ij}$'s are the eigenvalues; \{$|e_i\rangle^{\rm A}$\} and \{$|e_j\rangle^{\rm B}$\} are local eigenbases. A nonclassically correlated state $\rho_{\rm ncc}$  is a state which has no product eigenbasis.

There are measures of nonclassical correlation or quantumness
\cite{c-OZ01,c-OH02,c-H05,c-G07,c-SRN08,c-L08,c-DG09,c-SRN09-1,c-SRN09-2}
in several different contexts. Among these measures,
zero-way quantum deficit \cite{c-OH02, c-H05} is closely related to the
classical/nonclassical correlation description of our interest. It is a
nonlocalizable information under the zero-way CLOCC protocol
\cite{c-H05}, namely the protocol that allows two players to transfer
systems over a completely dephasing channel subsequent to local unitary
transformations. Nonclassical correlation is detected if the zero-way
quantum deficit is nonvanishing. Technically, the zero-way quantum deficit
is evaluated by minimization over all local bases of the discrepancy
between the von Neumann entropy of the state and the one after applying
the local unitary operations followed by local complete dephasing. 

For an experimental detection of nonclassical correlation, the currently
available measures are expensive because the complete information of
the density matrix of the system is required in general. This constraint may be accepted for a
quantification of nonclassical correlation; however, only for a
detection, it is too demanding.

We can use the approach that was taken 
in experimental detection of entanglement.
As the full state tomography is unnecessarily expensive, detection using entanglement witnesses
\cite{c-T02,c-B04} has become a widely-used approach.

Entanglement witness $W$ is an Hermitian observable with which one may
detect entanglement without state tomography \cite{c-DMK03}. It
satisfies that $\langle W\rangle \ge 0$ for any separable state, and
hence $\langle W\rangle <0$ implies the existence of entanglement. A
typical form of $W$ is $c - A$ with $c\ge 0$ and $A$ a certain Hermitian
positive matrix. In a simple case, $A = |\psi\rangle\langle\psi|$ with
$|\psi\rangle$ an entangled pure state. If $W$ is decomposed into experimentally feasible
observables then the detection can be done
without tomography. 

Because of the intrinsically operational benefits of using entanglement
witness, this concept has been not only developed extensively in its
theory \cite{M20, M21, M22, M23, M24}, but also in experiments
\cite{M25, M26, M27, M28, M29, M30}. 
The
approach of witness was taken also for the
Schmidt number \cite{SBL01} and the Slater number \cite{SCKLL01}. From
the operational side of interest, attempts have been concentrated on
decomposing the witness operators into local operations that are easily
measurable for a typical physical system of interest. 

It is important to implement a witness in a way that a single run
of an experiment is sufficient in the case where multiple copies of
a state is not easily obtained (e.g., when the generator and the
investigator are different). In \cite{ours}, such an implementation
has been introduced by using nondestructive ensemble average
measurements which enable simultaneous measurements of non-commuting
operators.

Our method to detect nonclassical correlation uses a map
working in a similar manner as entanglement witnesses but is workable
for detecting nonclassical correlation. We 
introduce a decomposition of a map in a way that a single run of
an experiment is sufficient to determine the outcome.

A map 
\[
 \mathcal{W}:\mathbbm{S}\rightarrow\mathbbm{R},
\]
with $\mathbbm{S}$ a state space, is called a nonclassical-correlation witness map if the following properties are satisfied.\\
(i) For any bipartite state $\rho_{\rm pcc}$ having a product eigenbasis, $\mathcal{W}\rho_{\rm pcc}\ge 0$ holds.\\
(ii) There exists a bipartite state $\rho_{\rm ncc}$ having no product eigenbasis such that $\mathcal{W}\rho_{\rm ncc}<0$.

A typical form of nonclassical-correlation witness maps is a map $\mathcal{W}$ in the form
\begin{equation}\label{eqTypical}
 \mathcal{W} \rho = c
- ({\rm Tr}\rho A_1)({\rm Tr}\rho A_2)\cdots({\rm Tr}\rho A_m)
\end{equation}
with $c\ge 0$ and $A_1,A_2,\cdots,A_m$ positive Hermitian matrices.

By construction, the nonclassical correlation witness of the above form is a nonlinear map. In general, to detect nonclassical correlation of
nonclassically-correlated separable states, {\it nonlinear} nonclassical correlation witness maps should be required. This is clear by the following theorem. 

\paragraph*{Theorem 1.} A linear nonclassical correlation witness map cannot detect nonclassical correlation of a separable state.
\paragraph*{Proof}
Let us write a linear nonclassical correlation witness map as $\mathcal{W}_{\rm lin}$. By the definition of a nonclassical correlation witness, for any properly-classically-correlated state $\rho_{\rm pcc}$, $\mathcal{W}_{\rm lin}\rho_{\rm pcc}\ge 0$ holds. Because of the linearity, for any convex combination of $\rho_{\rm pcc}$'s, namely, $\sum_{k}p_k\rho_{{\rm pcc},k}$ with nonnegative weights $p_k$ satisfying $\sum_k p_k=1$, we have $\mathcal{W}_{\rm lin} (\sum_{k}p_k\rho_{{\rm pcc},k})\ge0$. Recall that a separable state is a state in the form $\rho_{\rm sep} = \sum_k p_k |a_k\rangle\langle a_k|\otimes|b_k\rangle\langle b_k|$, with the same definition for $p_k$, where the projectors $|a_k\rangle\langle a_k|$'s ($|b_k\rangle\langle b_k|$'s) of the subsystem A (B) are possibly noncommutative. Obviously, $\rho_{\rm sep}$ is a convex combination of $\rho_{\rm pcc}$'s. Therefore $\mathcal{W}_{\rm lin}\rho_{\rm sep} \ge0$ holds.$\Box$

The nonclassical correlation witness of (\ref{eqTypical}) includes a typical form of entanglement witnesses. For example, the witness $W_{\rm Bell}=1/2-|{\rm Bell}\rangle\langle{\rm Bell}|$ with $|{\rm Bell}\rangle$ one of the two-qubit Bell states is equivalent to the nonclassical correlation witness map $\mathcal{W}_{\rm Bell}: \rho\mapsto 1/2 -{\rm Tr}\rho|{\rm Bell}\rangle\langle{\rm Bell}|$.

To construct a witness map in the form of (\ref{eqTypical}), an easy way is as follows. First we pick up eigenvectors $|v\rangle$ of some nonclassically correlated state and use $|v\rangle\langle v|$'s for $A_1, A_2,\ldots, A_m$ of RHS. (Of course, one may not follow this way in general.) Then we concentrate on choosing a proper constant $c$ of the RHS. The smallest possible $c$ is optimal in the detection range as is clear from the form. Since $c\ge ({\rm Tr}\rho A_1)({\rm Tr}\rho A_2)\cdots({\rm Tr}\rho A_m)$ should hold for any $\rho$ having a product eigenbasis, the optimal value for $c$ is the largest value of $({\rm Tr}\rho A_1)({\rm Tr}\rho A_2)\cdots({\rm Tr}\rho A_m)$ for $\rho$ having a product eigenbasis.

As an example, let us introduce the two-qubit state
\[
 \sigma=\frac{1}{2}(|00\rangle\langle00|+|1+\rangle\langle1+|)
\]
with $|+\rangle=\frac{|0\rangle+|1\rangle}{\sqrt{2}}$. It has no product eigenbasis because $|0\rangle\langle 0|$ and $|+\rangle\langle +|$ are non-orthogonal to each other. This state is an example of separable states having no product eigenbasis. Then, the following nonclassical correlation witness map is utilized.
\begin{equation}\label{eqW2}
 \mathcal{W}_\sigma: \rho\mapsto c - ({\rm Tr}\rho |00\rangle\langle00|)({\rm Tr}\rho|1+\rangle\langle1+|).
\end{equation}

The constant $c$ is chosen so that $\mathcal{W}_\sigma\rho \ge 0$ for any
$\rho$ having a product eigenbasis. To find the optimal constant
$c_{\rm opt}$, let us introduce the following lemma.
\paragraph*{Lemma 1.}
Let us set
$\tau = |s\rangle\langle s|\otimes \rho^{\rm B}$ 
where $|s\rangle = (|0\rangle + e^{i\theta}|1\rangle)/\sqrt{2}$
with $\theta$ any angle; $\rho^{\rm B}$ is a single-qubit state.
Then, $f(\rho_{\rm pcc})=
({\rm Tr}\rho_{\rm pcc} |00\rangle\langle00|)
({\rm Tr}\rho_{\rm pcc}|1+\rangle\langle1+|)$
is maximized for a state written in  the form of $\tau$.

The proof of Lemma 1 is given in Appendix 1. By Lemma 1, 
\[
 c_{\rm opt} = \underset{\tau}{\rm max} f(\tau)
=\underset{\rho^{\rm B}}{\rm max}\frac{1}{4}\langle 0|\rho^{\rm B}| 0\rangle
\langle +|\rho^{\rm B}| +\rangle.
\]
Let us set
\[
 \rho^{\rm B} = \begin{pmatrix}a & b \\b^* &1 - a\end{pmatrix}
\]
with $0\le a\le 1$ and $b\in\mathbbm{C}$.
Due to the positivity of $\rho^{\rm B}$, $|b|\le \sqrt{a(1-a)}$ holds.
Therefore we have
\[
 c_{\rm opt} = \underset{a,b}{\rm max}\frac{a[2{\rm Re}(b)+1]}{8}.
\]
The maximum value is obtained when $b=\sqrt{a(1-a)}$.
Thus
\[
 c_{\rm opt} = \underset{a}{\rm max}\frac{a[1+2\sqrt{a(1-a)}]}{8}.
\]
A straight-forward calculation suggests that the maximum value is
obtained when $a$ is $\hat a = \frac{2+\sqrt{2}}{4}$. Consequently, we have
\[
 c_{\rm opt} = \frac{\hat a[1+2\sqrt{\hat a(1-\hat a)}]}{8} = 0.182138\cdots.
\]

To see this works fine, for the state $\sigma$,
\[
 \mathcal{W}_\sigma\sigma=c_{\rm opt} - 0.250 < 0.
\]
Thus, a nonclassical correlation of $\sigma$ is successfully detected.

In a typical ensemble quantum information processing, nondestructive polarization measurements and global unitary operations are available in a single run experiment. Hence, it is possible to detect a polarization of a spin without demolition of a state. For such an ensemble system quantum information processing we prove that generally for an $n$-qubit system, a nonclassical-correlation witness map in the form (\ref{eqTypical}) can be decomposed in terms of global unitary operations and nondestructive polarization measurements on individual qubits.\\~

Consider the $i$th Hermitian operator $A_i$ of the $i$th factor of the second term of (\ref{eqTypical}). There exists a unitary transformation to diagonalize $A_i$. Thus, $A_i$ can be measured by finding populations of the density matrix after applying the unitary transformation to the given state. Each population is obtained by measuring the projector $|x_1\cdots x_n\rangle\langle x_1\cdots x_n|$ with $x_k\in\{0,1\}$. This projector is a tensor product of $n$ factors each of which is $\in (I\pm Z)/2$. Each term of this tensor product is thus a product of ($n-l$) $I$'s and $l$ $Z$'s ($0\le l \le n$). Such a term can be transformed into a single-qubit polarization, say, $Z\otimes I\otimes\cdots \otimes I$ by a certain global permutation operation, thus proving single-run-experiment detection of a nonclassical correlation of the form of (\ref{eqTypical}).

It is still nontrivial to find an effective decomposition for a given map. Let us give a simple example. Recall that any unitary transformation is trace preserving:
\[
 {\rm Tr}\rho M ={\rm Tr}U\rho U^\dagger UMU^\dagger = {\rm Tr}\hat\rho UMU^\dagger
\]
with $\hat\rho=U\rho U^\dagger$ and $M$ an observable. Using this fact, $\mathcal{W}_\sigma$ defined in (\ref{eqW2}) can be transformed in terms of local operations $I$ and $Z$ utilizing a global unitary operation $U$.

Let us set
\[
 U= {\rm controlled-}H=\begin{pmatrix}1&0&0&0\\
 0&1&0&0\\
0&0&1/\sqrt{2}&1/\sqrt{2}\\
0&0&1/\sqrt{2}&-1/\sqrt{2}
\end{pmatrix}.
\]
Then,
\[
|00\rangle\langle00|\overset{U}{\mapsto}|00\rangle\langle00| = \frac{1}{4}(1 + I\otimes Z + Z\otimes I +Z\otimes Z)
\]
and
\[
 |1+\rangle\langle1+|\overset{U}{\mapsto}|10\rangle\langle10|=\frac{1}{4}(1 + I\otimes Z - Z\otimes I - Z\otimes Z).
\]
We can measure the polarizations $<Z_1>={\rm Tr}\hat\rho(Z\otimes I)$ and $<Z_2>={\rm Tr}\hat\rho(I\otimes Z)$. These measurements can be non-demolition measurements in  e.g. NMR. We, however, need to have ${\rm Tr}\hat\rho(Z\otimes Z)$. We utilize the fact that 
\[
{\rm Tr}\hat\rho(Z\otimes Z)= {\rm Tr}{\rm CNOT}\hat\rho{\rm CNOT} (I\otimes Z).
\]
Thus the polarization of the second qubit for the state $\hat{\hat\rho}={\rm CNOT}\hat\rho{\rm CNOT}$ is equal to ${\rm Tr}\hat\rho(Z\otimes Z)$.

\begin{figure}
\begin{center}
 \scalebox{0.9}{\includegraphics{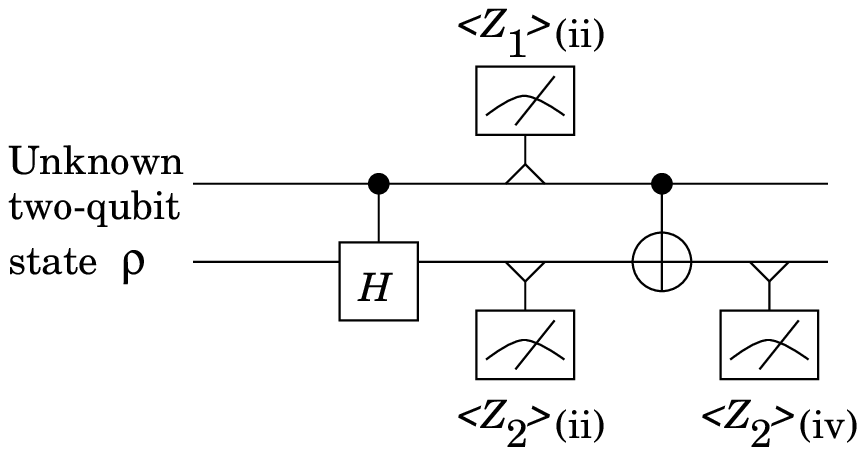}}
\caption{\label{figex2} Quantum circuit to achieve enough
data for calculating $\mathcal{W}_\sigma\rho$. Three nondestructive
ensemble measurements are used.}
\end{center}
\end{figure}

Thus, what we should do to evaluate $\mathcal{W}_\sigma\rho$ is firstly to apply ${\rm controlled-}H$ to map $\rho$ to $\hat\rho$. Then, we should measure the polarizations of the qubits, $<Z_1>_{\rm (ii)}$ and $<Z_2>_{\rm (ii)}$. After that, we apply ${\rm CNOT}$ to map $\hat\rho$ to $\hat{\hat\rho}$ and finally measure the polarization of the second qubit, $<Z_2>_{\rm (iv)}$. These steps are illustrated in Figure \ref{figex2}. The value of $\mathcal{W}_\sigma\rho$ is calculated as
\begin{eqnarray}
\mathcal{W}_\sigma\rho&=&c-\frac{1}{16}(1+<Z_1>_{\rm (ii)}+<Z_2>_{\rm (ii)}+<Z_2>_{\rm (iv)}) \nonumber\\ \nonumber
&{}&\times (1-<Z_1>_{\rm (ii)}+<Z_2>_{\rm (ii)}-<Z_2>_{\rm (iv)}).
\end{eqnarray}

In summary, for the concept of nonclassical correlation defined
differently from entanglement, we have introduced the witness maps. These
maps have been shown to be useful for detecting nonclassical correlation
without expensive state tomography. In order to construct a witness map,
a constant $c$ should be determined appropriately. We have seen an example
for which the optimal value for $c$ has been found analytically. However,
the analytical approach to find optimal $c$ is not easy in general. For
such a case, it can be estimated by a numerical search.  An example
is given in Appendix 2. We have also investigated a way to implement
a witness map in a single-run implementation in bulk-ensemble systems,
such as NMR and ENDOR. A single-run implementation is necessary in case
multiple copies of a state are not available. The implementation we
have proposed takes advantage of nondestructive polarization measurements
equipped in the bulk-ensemble systems.

R.R. is supported by the Japan Society for the Promotion of Science (JSPS) through its ``Funding Program for World-Leading Innovative R\&D on Science and Technology (FIRST Program)." R.R. and A.S. are supported by Grant-in-Aids for Scientific Research from JSPS (Grant Nos. 1907329 and 21800077, respectively). A.S. is supported by the ``Open Research Center'' Project for Private Universities: matching fund subsidy from MEXT.

\paragraph*{{\bf Appendix 1:}} Proof of Lemma 1.
\begin{proof}
A two-qubit state having a product eigenbasis (PE) is in general
\[
 \rho_{\rm PE}=\sum_{i,j=0}^1e_{ij}|u_i\rangle\langle u_i|
\otimes |v_j\rangle\langle v_j|
\]
with the $ij$'th eigenvalue $e_{ij}$ and the corresponding eigenvector
$|u_i\rangle|v_j\rangle$.
Without losing the generality, we assume that
$e_{00} \ge e_{01} \ge e_{10} \ge e_{11}$.
We have
\[
 f(\rho_{\rm PE})= \sum_{ijkl}e_{ij}e_{kl}
|\langle 0|u_i\rangle|^2|\langle 0|v_j\rangle|^2
|\langle 1|u_k\rangle|^2|\langle +|v_l\rangle|^2.
\]
We rewrite this as
\[
\begin{split}
&f(\rho_{\rm PE})=\sum_{jl}|\langle 0|v_j\rangle|^2|
\langle+|v_l\rangle|^2\bigl(e_{0l}|\langle 0|u_1\rangle|^2\\
&+e_{1l}|\langle 0|u_0\rangle|^2\bigr)
\bigl(e_{0j}|\langle 0|u_0\rangle|^2+e_{1j}|\langle 0|u_1\rangle|^2\bigr).
\end{split}
\]
Let us write $p=|\langle 0|u_0\rangle|^2$. Then we have
$|\langle 0|u_1\rangle|^2 = 1-p$.
This leads to
\[
\begin{split}
 &f(\rho_{\rm PE})=\sum_{jl}|\langle 0|v_j\rangle|^2|
\langle+|v_l\rangle|^2
\bigl[e_{0j}e_{0l}p(1-p)\\
& + e_{0j}e_{1l}(1-p)^2 + e_{0l}e_{1j}p^2 + e_{1j}e_{1l}p(1-p)\bigr].
\end{split}
\]
There are two cases: (i) $p(1-p)$ is the largest among $p^2, p(1-p),$ and
$(1-p)^2$. (ii) either $p^2$ or $(1-p)^2$ is the largest among the factors.\\
~~Case (i): As $p(1-p)$ is the largest among the factors involving $p$,
the largest value of $f(\rho_{\rm PE})$
is obtained when the weights are shifted to 
$e_{0j}e_{0l}+e_{1l}e_{1j}$ with $e_{0j}e_{1l}$ and $e_{0l}e_{1j}$
set to zero. This implies that $e_{1j}=e_{1l}=0$. Therefore,
$\rho_{\rm PE}$ with the subsystem $A$ in a pure state
makes $f(\rho_{\rm PE})$ the largest.\\
~~Case (ii): As $p^2$ or $(1-p)^2$ is the largest among the factors
involving $p$, the largest value of $f(\rho_{\rm PE})$ is obtained
for the weights $e_{00}=e_{01}=e_{10}=1/3$ and $e_{11}=0$. This is
because of the order of the eigenvalues we assumed.
Note that $p^2+(1-p)^2\le 1$ and $p(1-p)\le 1/4$.
The largest value of $f(\rho_{\rm PE})$ is then bounded above by
$\sum_{jl}|\langle 0|v_j\rangle|^2|\langle+|v_l\rangle|^2
(1/36+1/9+1/36)=1/6=0.1666\cdots$.
This is less than the largest value for case (i) that turns out
to be $0.1821\cdots$.
\end{proof}
\paragraph*{{\bf Appendix 2:}} Numerical example.~\\
Consider the witness map $\mathcal{W}_{02+}$
acting on a two-qutrit density matrix $\rho$, defined as
\[
 \mathcal{W}_{02+}:\rho\mapsto
c-({\rm Tr}\rho A_1)({\rm Tr}\rho A_2)({\rm Tr}\rho A_3)
\]
with $A_1=|02\rangle\langle02|$, $A_2=|+0\rangle\langle+0|$, 
$A_3=|2+\rangle\langle2+|$, and $|+\rangle=\frac{|0\rangle+|1\rangle}{\sqrt{2}}$.
The optimal $c$ is given by
\[
 \underset{\rho_{\rm pcc}}{\rm max}~
({\rm Tr}\rho_{\rm pcc} A_1)({\rm Tr}\rho_{\rm pcc} A_2)({\rm Tr}\rho_{\rm
pcc} A_3).
\]
After a search over $6.4\times 10^9$ $\rho_{\rm pcc}$'s generated
randomly, the value $0.019897$ has been found. (A state
with purity $0.51664$ and four distinct nonzero eigenvalues
gave this value.) From this numerical result, we conjecture
that the exactly optimal $c$ is less than $\tilde c=0.02$.
The witness map with $\tilde c$ for $c$ is usable for detecting
nonclassical correlation of the state
$\rho_{02+}=\frac{1}{3}(A_1+A_2+A_3)$; we have
$\mathcal{W}_{02+}\rho_{02+}=\tilde c - 1/27 < 0$.

\end{document}